\documentclass[10pt,conference]{IEEEtran}
\usepackage[compatibility=false]{caption}
\newif\ifDEBUG
\DEBUGfalse
 \DEBUGtrue
\newif\ifANONYMOUS
\ANONYMOUSfalse

\usepackage{cite}
\usepackage{arydshln} 
\usepackage{graphicx}
\usepackage{textcomp}
\usepackage{caption}
 \usepackage{tikz}
\usetikzlibrary{arrows.meta,positioning,fit,backgrounds,shapes,calc}
\usepackage[dvipsnames]{xcolor}
\def\BibTeX{{\rm B\kern-.05em{\sc i\kern-.025em b}\kern-.08em
    T\kern-.1667em\lower.7ex\hbox{E}\kern-.125emX}}

\usepackage{amsmath,amsfonts}
\usepackage{algorithmic}
\usepackage{graphicx}
\usepackage{textcomp}
\usepackage{xcolor}
\usepackage{booktabs} 
\usepackage{xspace} 
\usepackage[normalem]{ulem}
\usepackage{makecell}
\usepackage{tcolorbox}
\usepackage{enumitem}
\usepackage{siunitx}
\usepackage[vskip=1em,font=itshape,leftmargin=2em,rightmargin=2em]{quoting}
\usepackage{svg}
\usepackage{listings}
\usepackage{subfigure}
\usepackage{xcolor}
\PassOptionsToPackage{hyphens}{url}\usepackage{hyperref}

\usepackage{soul}
\usepackage{hyperref}
\usepackage{cleveref}

\ifDEBUG
   \newcommand{\JD}[1]{\textcolor{purple}{[Jamie says: #1]}}
    \newcommand{\KC}[1]{\textcolor{blue}{[Kelechi says: #1]}}
    \newcommand{\TRS}[1]{\textcolor{olive}{[Taylor says: #1]}}
    
    \newcommand{\SA}[1]{\textcolor{orange}{[Santiago says: #1]}}
    \newcommand{\SO}[1]{\textcolor{pink}{[Sofia says: #1]}}
    
\else
    \newcommand{\JD}[1]{}
    \newcommand{\KC}[1]{}
    \newcommand{\TRS}[1]{}
    \newcommand{\AK}[1]{}
    \newcommand{\SA}[1]{}
    \newcommand{\SO}[1]{}
    
\fi

\usepackage{balance} 

\usepackage{amsmath}
\usepackage{cleveref}

\crefformat{section}{\S#2#1#3}
\crefname{figure}{Figure}{Figures}
\crefname{appendix}{Appendix}{Appendices}
\crefname{table}{Table}{Tables}
\crefname{algorithm}{Algorithm}{Algorithms}
\crefname{listing}{Listing}{Listings}
\crefname{theorem}{Theorem}{Theorems}
\crefname{thm}{Theorem}{Theorems}
\crefname{lemma}{Lemma}{Lemmata}
\crefname{equation}{Eqt.}{Eqts.}
\crefformat{Grammar}{Grammar #1}

\usepackage{enumitem}
\usepackage{booktabs}

\usepackage{makecell}
\usepackage{multirow}
\usepackage[T1]{fontenc}

\newcommand{\eg}{\textit{e.g.,} }
\newcommand{\etal}{\textit{et al.}\xspace}

\usepackage{enumitem}

\usepackage{quoting}

\DeclareUnicodeCharacter{2060}{\nolinebreak}

\newcounter{finding}

\begin{document}

\title{Why Software Signing Matters: A Threat Dimension and Boundary Perspective}
\title{Why Software Signing (Still) Matters: \\ Trust Boundaries in the Software Supply Chain}

\ifANONYMOUS
    \author{Anonymous author(s)}
    
\else

\author{
    \IEEEauthorblockN{Kelechi G. Kalu}
    \IEEEauthorblockA{Purdue University\\
    kalu@purdue.edu}
    \and
    \IEEEauthorblockN{James C. Davis}
    \IEEEauthorblockA{Purdue University\\
    davisjam@purdue.edu}
}

\fi

\maketitle


\begin{abstract}
Software signing provides a formal mechanism for provenance by ensuring artifact integrity and verifying producer identity.
It also imposes tooling and operational costs to implement in practice.
In an era of centralized registries such as PyPI, npm, Maven Central, and Hugging Face, it is reasonable to ask whether hardening registry security controls obviates the need for end-to-end artifact signing.
In this work, we posit that the core guarantees of signing, provenance, integrity, and accountability are not automatically carried across different software distribution boundaries. These boundaries include mirrors, corporate proxies, re-hosting, and air-gapped transfers, where registry security controls alone cannot provide sufficient assurance.
We synthesize historical practice and present a trust model for modern distribution modes to identify when signing is necessary to extend trust beyond registry control.
Treating signing as a baseline layer of defense strengthens software supply-chain assurance even when registries are secure.

\end{abstract}

\section{Introduction}

The software supply chain involves the creation and distribution of software packages—from their point of origin, through intermediate distribution channels, and finally to their destinations where they are installed and executed.
Security in this model focuses on two core aspects: ensuring the \textit{integrity} of the software artifacts and verifying the \textit{authenticity} of their origin.
These two properties determine whether downstream consumers can trust the integrated software component.

Prior work has proposed various mechanisms to achieve these security goals.
Most notably, software signing, defined as the cryptographic association of an identity with a software artifact, has been widely recommended to verify both artifact integrity and producer authenticity.
However, multiple empirical studies~\cite{usenix_2025_signing_interview_kalu, Schorlemmer_Kalu_Chigges_Ko_Ishgair_Bagchi_Torres-Arias_Davis_2024} have found inconsistent adoption, especially across open-source ecosystems.
Signing is cross-cutting: it requires the creation of signatures by producers and their verification by consumers or intermediaries at multiple points in the software lifecycle.
This distributed requirement, coupled with limited incentives and uneven infrastructure\cite{kalu2025johnnysignsnextgenerationtools}, makes effective deployment difficult in practice.

These challenges have led some to question whether the effort required to implement signing is worth the cost, particularly when centralized software registries such as PyPI, npm, and Maven Central are increasingly being hardened.
These registries serve as critical distribution chokepoints and already implement various protections, such as publisher authentication, audit logging, and access control.
Because they operate at scale, hardening these points may seem like a more straightforward and scalable approach to improving software supply chain security.
If registries are trusted, why not rely on checksums?
Checksums verify content integrity but not origin or publisher identity, so they do not provide actor accountability.
However, a linear view of software distribution as ``Upstream $\rightarrow$ Registry $\rightarrow$ Downstream user'' is overly simplistic and does not reflect modern distribution.

We argue that software signing extends trust across boundaries that registries do not, or cannot, control.
We treat signing as a baseline layer of defense that enables verifiable trust across decentralized and evolving distribution paths.
We revisit the role of signing not as a fallback or redundant measure, but as a foundational mechanism for ensuring verifiable trust across complex, decentralized, and evolving distribution paths.
Through our analysis, we show why software signing continues to matter, even in a world of increasingly secure registries.

In summary, this position piece frames software signing as a form of portable trust and makes the following contributions:
\begin{itemize}
    \item \textbf{Position:}
    We provide a historically grounded perspective on why software signing arose and remains necessary in modern software distribution channels.
    We clarify how signing complements registry hardening rather than competing with it.
    \item \textbf{Trust model and analysis:}
    We present a threat and trust model across common distribution modes and show how guarantees change as artifacts cross trust boundaries.
    We address misconceptions, showing that hardened registries do \textit{not} render signing redundant.
\end{itemize}

\section{Background}

We first describe the security properties of software supply chains that warrant the use of software signing (\cref{sec:bg-SSC_Properties}). We then provide an overview of software signing and its relationship to these properties (\cref{sec:bg-signing}), and finally, we review the historical evolution of software distribution models (\cref{sec:bg-distribution-history}).

\subsection{Software Supply Chains and Their Security Properties}
\label{sec:bg-SSC_Properties}



When a modern software application executes, the code within it comes from both the provider's software engineers and the components that they integrated.
The integrated components may be written by other teams within the same organization (\eg an internal team that creates utility routines or device drivers), from open-source projects (\eg the React framework\cite{react_website}), or from third-party vendors (\eg a real-time operating system such as SafeRTOS\cite{highintegritysystemsSAFERTOSSafety}).
Those components may themselves depend on other components, producing a dependency graph also called the \textit{software supply chain}\cite{singla_empirical_2023}.
Measurements by Synopsys suggest that the bulk of code in modern applications is integrated rather than custom. Specifically, their analysis showed that open-source components now make up about 80\% of code in typical codebases~\cite{synopsys_ossra_2023}.
To have confidence in the resulting application, the provider's engineering team must vet each component.
When considering their vetting process, they must keep their desired security properties in mind.

For any software application, engineers must consider the standard security properties of confidentiality, integrity, and availability, the ``CIA Triad''\cite{samonas2014cia}.
Those properties consider how the software application interacts with its deployment environment.
Each of those properties may be influenced by the integrated components, resulting in security properties that software engineers must consider that are specific to the software supply chain.
Following the analysis of Okafor \etal\cite{okafor_sok_2022}, the supply chain security properties are:

\begin{enumerate}
    \item \textbf{Transparency:} All actors in the supply chain should have access to information about the entire chain, enabling them to assess risks and apply countermeasures as needed. 
    \item \textbf{Validity:} The supply chain must maintain the integrity of its operations, artifacts, and actors. 
    \item \textbf{Separation:} The supply chain should minimize unnecessary connections and keep logically distinct operations, artifacts, and actors separate. 
\end{enumerate}

These security properties define the core design philosophies underlying many of the techniques developed to safeguard software production and distribution. 
For example, virtualization and containerization enforce the principle of \textit{separation} by compartmentalizing execution environments and reducing the attack surface. 
Software Bills of Materials (SBOMs) promote \textit{transparency} by making the composition of artifacts and operations visible. Multi-factor authentication enforces \textit{validity} by strengthening assurances around the actors who participate in the supply chain. 
Some techniques provide guarantees that span multiple properties. 
For instance, code scanning promotes both \textit{transparency} and \textit{validity} by exposing vulnerabilities in artifacts and confirming their integrity, though such measures remain limited to artifact security. 
A mechanism that directly connects artifact integrity to publisher authenticity is \textit{software signing}.
This technique not only reinforces the properties outlined above but also provides end-to-end assurances that extend across otherwise fragmented parts of the supply chain, making it a cornerstone of modern supply chain security.

\subsection{Software Signing}
\label{sec:bg-signing}

\begin{figure}[htbp]
    \centering
    \includegraphics[width=0.95\linewidth]{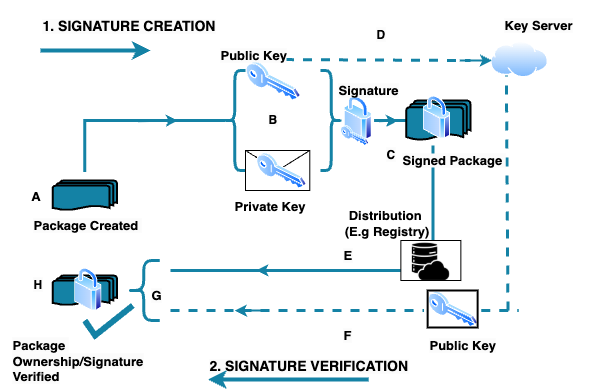}
    \caption{ 
    \textbf{Typical workflow for software signing}.
    The component author packages (A) and signs (B) their software.
    The signed package (C) and public key (D) are published.
    To use a package, a user downloads it (E) and its public key (F) and verifies the signature (G).
    Figure adapted from Kalu et al.~\cite{usenix_2025_signing_interview_kalu}.
    }
    \label{fig:Signing-fig}
\end{figure}

Software signing (or code signing) is a cryptographic process that binds a verifiable producer identity to a specific software artifact. At its core, a producer computes a cryptographic digest of the artifact and attests to it, typically by creating a digital signature with a private key, so that consumers can later verify, using the corresponding public key, that the artifact originated from the claimed publisher and has not been altered since attestation. 
In this work, we use “software signing” as an \emph{umbrella term} for mechanisms that provide the same verifiable binding of \emph{identity} to \emph{content}, including detached signatures, signed release manifests (e.g., \texttt{SHA256SUMS.sig}), package-manager metadata signatures, transparency-log attestations, and publisher-issued checksum digests distributed via authenticated, publisher-controlled channels. These assurances address alteration attacks such as unauthorized code injection, dependency/package confusion, and distribution of malicious versions~\cite{ladisa_sok_2023}.

By combining integrity (the artifact has not changed) with authenticity (it was produced by the claimed publisher), signing yields signature-backed provenance of origin and content. In supply-chain terms, this enhances \emph{transparency} by making identity and chain-of-custody verifiable, strengthens \emph{validity} by authenticating publishers and artifacts, and supports \emph{separation} by reducing reliance on centralized intermediaries. Consistent application also aligns with \emph{Zero Trust} principles, \emph{never trust, always verify}\cite{stafford2020zero}, by enabling consumers and automation to validate artifacts independently of network location or intermediary reputation. 


Historically, software signing first emerged in the context of operating systems, where platform vendors introduced code-signing mechanisms to prevent the execution of untrusted binaries. For example, Microsoft’s Authenticode was publicly announced in June 1996 alongside Internet Explorer 3.0 and ActiveX, enabling digital signatures on executables and controls to verify origin and integrity~\cite{microsoft_authenticode_1996}. In the open-source ecosystem, Linux distributions such as Debian integrated similar signing into their package workflows: GPG is used to sign and verify repository metadata (\eg the \texttt{Release} file), ensuring that packages installed via \texttt{apt} can be traced back to trusted maintainers~\cite{debian_secureapt_manual, debian_secureapt_wiki}. These early adoptions laid the foundation for signing in more complex distribution contexts, first at the operating system level\cite{apple_distribution_signed_mac, microsoft_authenticode_learn}, then in package managers, and eventually in modern registries\cite{sonatypeWorkingWith, dockerContentTrust}.
This progression illustrates how software signing evolved in response to shifts in distribution practices, from OS-level trust anchors, to package managers, to registries that now serve millions of developers. Because the evolution of signing cannot be separated from the evolution of distribution itself, the next section turns to the history of software distribution, tracing the shift from isolated in-house development, to early peer-to-peer code-sharing platforms, to today’s centralized registries.

\subsection{Software Distribution}
\label{sec:bg-distribution-history}



\begin{figure*}[htbp]
    \centering
    \includegraphics[width=0.95\textwidth]
    {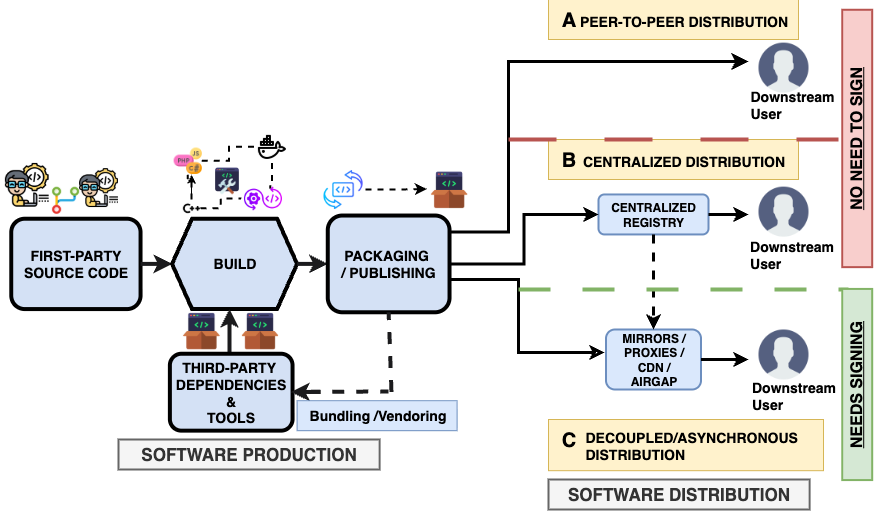}
    \caption{\textbf{Evolution of software production and distribution.} 
The left side shows a software-factory pipeline\cite{slsaSupplychainLevels}, first-party source, third-party dependencies/tools, build, and packaging/publishing, with optional \emph{bundling/vendoring} of dependencies at release. 
The right side illustrates distribution paths: historically, project-specific or peer-sharing hubs (Top); modern centralized registries (\eg PyPI\cite{pypi}, npm\cite{npmjsHome}, Maven Central\cite{maven_central}) that improve access control and metadata integrity but introduce single points of control/failure (Middle); and redistribution beyond the registry via mirrors, enterprise proxies/caches, CDNs, or air-gapped transfer (Bottom). 
When artifacts traverse beyond the registry’s direct control, \emph{software signing} is needed to carry integrity and authenticity across those boundaries. 
\emph{The figure distinguishes pre-publish steps (\eg Bundling and Vendoring) in the software factory, that may influence the need/lack of need for signing, from post-publish delivery via the registry and decentralized post-publish redistribution (mirrors, proxies, air-gaps).}}

    \label{fig:distribution-model}
\end{figure*}

The need to consider security properties in the supply chain has changed over time alongside the engineering processes of software organizations. Historically, software was often developed in isolation, with little to no role for external components. Although software engineering visionaries proposed reusable libraries and components\cite{mcilroy68b}, for example, through the object-oriented design movement, the industry did not begin to widely adopt them until the 1990s\cite{booch1994ooa}. Even then, components were often developed and used by the same organization that used them, rather than integrating third-party code. 
This landscape began to shift in the late 1990s and early 2000s with the growth of public code-sharing platforms\cite{Sourceforge}. Early platforms such as SourceForge provided centralized hosting and distribution for open-source projects~\cite{tamburri2020canary, Sourceforge}. These were soon followed by platforms like GitHub and GitLab, which combined distributed version control with social coding features~\cite{dabbish2012social}, dramatically accelerating collaboration across organizational boundaries. 

During this same period, government and industry attitudes toward open-source software also evolved. The U.S. Department of Defense (DoD), which for decades viewed OSS with skepticism, formally revised its stance in the 2009 Wennergren memo~\cite{wennergren2009clarifying}, recognizing OSS as commercial software and encouraging its use under appropriate conditions. More recently, DoD initiatives such as Platform One\cite{DoDPlatformOne} and the Iron Bank container registry\cite{DoDIronBank} have actively promoted the integration of OSS into defense systems.

As open-source adoption grew, centralized language-specific registries emerged to manage the distribution of third-party dependencies. Registries such as PyPI for Python, npm for JavaScript, and Maven Central for Java became indispensable for modern software development.
PyPI alone indexes on the order of 670k+ packages~\cite{PSF}; the npm ecosystem contains millions of packages and serves tens of billions of weekly downloads~\cite{Pinckney_Cassano_Guha_Bell_2023}; Maven Central has grown into one of the largest Java component repositories, with sustained growth documented in recent empirical studies~\cite{Ede_Dietrich_Zülicke_2025}; and the Hugging Face registry has emerged as a central registry for machine-learning models and datasets used across industry and academia~\cite{hf_hub_docs}. Importantly, a significant fraction of these downloads is machine-initiated (e.g., CI/CD pipelines and automated rebuilds), which magnifies both the operational benefits and the potential blast radius of supply-chain incidents~\cite{zlatanB_2025}. 
These registries automated dependency management and lowered the barrier to reuse, but they also concentrated risk: a single compromised package or maintainer account could cascade into vulnerabilities affecting thousands or even millions of downstream systems.


These ecosystem changes reshaped not only who produces and hosts software, but also how(the manner) artifacts are packaged and delivered, evolving from simple project-host downloads in the SourceForge era to registry-mediated delivery and multi-hop redistribution via mirrors, enterprise proxies, CDNs, and air-gapped transfers, setting the stage for the more complex distribution paths discussed next.

\subsubsection{From Straightforward Downloads to Complex Distribution Paths}

As ecosystems grew, the path from producer to consumer diversified beyond a simple “download from the registry.” Scale, organizational boundaries, and performance/availability needs introduced additional actors and hops in the distribution pipeline.

\paragraph{Bundling \& Vendoring}
Producers may ship a single, self-contained deliverable that embeds dependencies to simplify downstream use. 
\emph{Examples:} Java \emph{uberjars}/fat JARs that include transitive JARs (Maven/Gradle)\cite{Baeldung_2017_fatjars}; statically linked binaries in Linux distributions\cite{Varbhat_2020_statically_linked}; container images that vendor OS and language libraries\cite{fatherlinux_2020_container_os}; front-end bundles that inline third-party modules\cite{Fateh_2025_webpack, Rem_2023_frontend}.

\paragraph{Proxies and enterprise caches}
Organizations often route dependency requests through repository managers or caching proxies that front public registries. 
\emph{Examples:} npm/PyPI corporate proxies in CI\cite{ProxyWing_2025}; Maven “remote repository” caches in Nexus/Artifactory\cite{rathore_nexus_linkedin_pulse}; internal Docker registries that mirror Docker Hub images\cite{renovate_discussion_internal_registry}.

\paragraph{Public mirrors and CDNs}
Ecosystems commonly replicate artifacts across geographically distributed mirrors and CDN edges to improve availability and throughput. 
\emph{Examples:} Debian/Ubuntu mirror networks for APT\cite{Debian_2025_mirror}; CPAN mirrors for Perl modules\cite{cpan_sites}; historical PyPI mirrors\cite{StableBuild-Docs_2024_pypi-mirror}; Maven Central CDN edges\cite{Central.Sonatype_2021}; Hugging Face mirrors for models and datasets\cite{MindNLP_2024_huggingface}.

\paragraph{Air-gapped transfer}
In regulated or sensitive environments, artifacts are exported from a connected network and imported into an air-gapped enclave via removable media or controlled gateways. 
\emph{Examples:} Exporting Maven/PyPI/conda repositories to removable media (\textit{sneakernet}\cite{jaya_sneakernet}); importing container images into offline enclaves\cite{nergaard_offline_docker_image}.





Taken together, this historical arc, from isolated organizational development, to early code-sharing platforms, to today’s centralized registries and other complex distribution paths, illustrates how the attack surface of software distribution has steadily expanded. The importance of software signing becomes more pronounced when we consider the underlying trust assumptions in these systems. While centralized registries such as PyPI, Maven Central, and Docker Hub enforce varying levels of access control and metadata integrity, they also act as single points of control, and potentially, single points of failure. As distribution workflows grow more complex and span multiple organizational, infrastructural, or technical domains (\eg through mirrors, corporate proxies, or offline transfers), trust must be re-established or extended across each of these boundaries.

\noindent
\subsection{\textit{Threat Model -- Base Assumptions.}}
\label{sec: threat-model}

For this position paper, we make the following assumptions:

\begin{enumerate}
    \item We assume that trusted and secure communication channels exist between entities that directly exchange software artifacts or components.
    \begin{enumerate}
        \item We assume that Peer-to-Peer distribution is between two established, trusted, connected entities.
    \end{enumerate}
    
    \item We assume that for centralized registries, the registry maintains a secure connection with producer entities, and that no man-in-the-middle attack is possible between producer and registry.
    \begin{enumerate}
        \item We assume that connections used by downstream consumers to obtain software components from registries are always secure.
        \item We assume that mirrors, caches, and airgapped environments cannot be fully trusted and may be insecure~\cite{cappos2008look}.
    \end{enumerate}
\end{enumerate}

Next, we introduce our framework for reasoning about these dynamics through the lens of trust boundaries and trust dimensions in software distribution.

\section{Trust Dimensions and Boundaries in Software Distribution}
\label{sec: Trust_Dimmension_and_Boundaries}



\begin{table*}[t]
\centering
\caption{
\textbf{Summary of trust dimensions by distribution model.}
We summarize the number of trust boundaries, defined by the number of entities, whether registry-local controls travel with artifacts, whether the model concentrates a single point of control or failure, and whether software signing is needed.
This table guides when registry hardening suffices and when portable verification via signing is necessary.
}
\label{tab:trust-dimensions}
\begin{tabular}{p{0.34\linewidth}p{0.12\linewidth}p{0.18\linewidth}p{0.16\linewidth}p{0.16\linewidth}}
\toprule
\textbf{Distribution model} & \textbf{Entities ($n$)} & \textbf{Single point of control/failure?} & \textbf{Need for signing} \\
\toprule
Peer-to-peer or project hubs & $n=2$ & No & None$\rightarrow$Optional \\
Centralized registries & $n=3$ & Yes & Optional$\rightarrow$Important \\
Decoupled/asynchronous (mirrors, CDNs, proxies) & $n=4$ & No & Important \\
\bottomrule
\end{tabular}
\label{tab:summary-dimmension}
\end{table*}

A central idea in our analysis of software signing is the role of trust in software distribution systems. While many package ecosystems are mediated by centralized registries (\eg PyPI, Maven Central), these systems may often rely on broader distribution infrastructures that introduce additional entities, like those described in \cref{sec:bg-distribution-history}.
Each such entity introduces a new trust boundary: a point where the recipient must trust that this entity, which they may not control, has preserved the integrity and authenticity of the software.

As verification must span more independent entities in the delivery path, the number of trust relationships and boundaries, \textit{trust surface}, increases\cite{manadhata2010attack, stafford2020zero}. Correspondingly, the need for end-to-end integrity mechanisms such as software signing grows. We therefore classify software delivery and distribution by the number of trust boundaries: the more entities in the chain, the less one can rely on implicit or registry-based integrity (illustrated in \cref{fig:distribution-model} and summarized in \cref{tab:summary-dimmension}).
\begin{enumerate}
    \item \textbf{Peer-to-Peer Distribution}: Packages are moved directly from one entity to another entity.
    In this distribution setup, there is typically a high level of trust between the entities involved in the transfer, and in most scenarios the parties have established a previous connection.
    The security of this process depends on the two engaging entities in this distribution, so the trust surface involves two(2) entities (n=2), the software producer and the downstream user.
    
    \item \textbf{Centralized Distribution}: Centralized distribution follows situations where there are established central registries or platforms where software components are centrally stored and then distributed to downstream entities (\eg npm, Rust's Cargo, Maven Central).
    The trust surface of this model is defined by the interaction of three(3) entities (n=3)—the software producer, the registry and the downstream user.
    This is an increased trust surface from the earlier peer-to-peer model, and thus there are more opportunities for alteration attacks: unauthorized publication or modification at the registry (compromised maintainer accounts or operator pipelines), tampering-in-transit and mirror substitution, cache poisoning, and replay/downgrade attacks that serve stale or forged packages and metadata, between the links of the three entities

    In recent years, registries have added controls, \eg publisher authentication, MFA for publishing, audit logs, and signed metadata, that strengthen assurances for consumers about who published a package and what they are downloading at the moment of retrieval.
    We analyze when registry hardening renders signing redundant \cref{sec:analysis-by-model}.

    \item \textbf{Decoupled/Asynchronous Distribution}: Situations where packages are mirrored, cached, delivered via CDNs and proxies, or transferred outside real-time registry access.
    In cases like these, software components originally distributed via centralized registries may also be distributed via additional redundant sources to ease network traffic or improve availability.
    The versions served by these replicas may fall out of sync with the primary registry.
    In this case, the trust surface becomes n=4, with four different entities involved in establishing trust in the distribution pipeline of these components (the software producer, the registry, mirrors/CDN, etc, and the downstream user).
    The class of attacks that can affect a distribution arrangement like this has already been discussed by Cappos \etal~\cite{cappos2008look}.
    Additionally, internal mirrors increase the trust surface(\eg via insider threats).
    Another notable context is air-gapped or military environments, where artifacts are exported from connected networks and imported offline into isolated enclaves.
    This further decouples distribution from registry policy and audit logs, and heightens the need for portable, offline verification.
    
\end{enumerate}

This distribution trust model framework allows us to reason about when and why software signing becomes not just beneficial, but essential.

We first discuss what hardening the security of distribution registries entails, and then describe cases where software signing remains essential.

\subsection{Analyzing guarantees by distribution model: when is signing necessary?}
\label{sec:analysis-by-model}

\begin{figure*}[ht]
  \centering
  \includegraphics[width=0.49\textwidth]{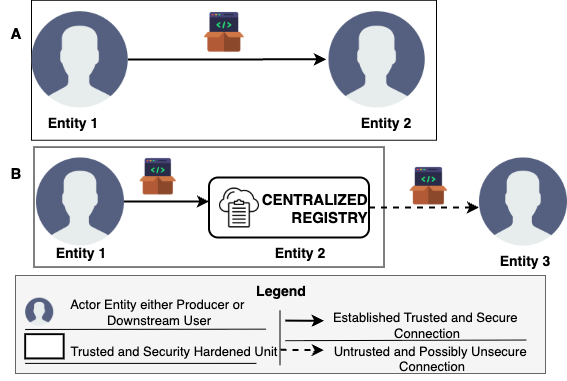}%
  \hfill
  \includegraphics[width=0.49\textwidth]{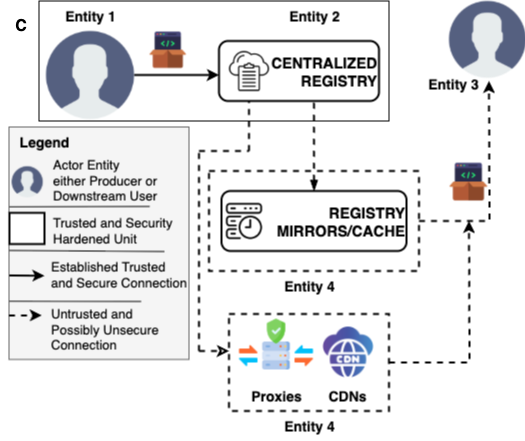}%
  \caption{\textbf{Trust Consideration in a Peer-to-Peer, a Centralized, and an Asynchronous Distribution Model.} 
A. \textit{Peer-to-Peer Distribution.} Producers (Entity~1) establish connections and directly distribute software components to downstream users.
B. \textit{Centralized Distribution.} Producers (Entity~1) publish artifacts to a trusted and security-hardened registry (Entity~2), from which downstream users obtain software components. 
This model improves access control and metadata integrity compared to peer-to-peer sharing, but it also concentrates risk in a single point of control or potential failure.
  C. Artifacts published to a centralized registry are replicated through secondary infrastructure such as mirrors and caching proxies. This model improves scalability and performance but introduces additional trust boundaries, since distribution may occur outside the direct control of the registry. }
  \label{fig:heatmapsRQ1}
\end{figure*}

We assess the strongest guarantees each distribution model can offer under ideal hardening, because decisions about where to invest controls (registry hardening vs.\ portable verification) depend on those limits. We then examine whether these guarantees obviate or still require software signing.


\paragraph{Peer-to-peer or project hubs}
Where two entities have an established trust relationship and exchange artifacts over authenticated, confidential channels, additional code signing may be optional for routine transfers.
This assumes a stable bilateral relationship, no redistribution beyond the pair, and no need for portable audit trails.
Signing becomes valuable when artifacts are archived for later verification, redistributed to third parties, or when personnel, infrastructure, or expectations change.

\paragraph{Centralized registries}
Within a hardened registry boundary, publisher authentication, audit logging, TLS, authorization, and policy, signing can be redundant for in-boundary fetches because integrity and identity are mediated by the registry at time of download.
However, registry-local controls do not travel with artifacts, and network or mirror-level attacks have historically subverted distribution paths\cite{cappos2008look}.
Therefore, signing remains necessary whenever trust must extend beyond the registry boundary or persist over time.

\paragraph{Decoupled or asynchronous delivery (mirrors, CDNs, proxies, caches)}
These infrastructures introduce additional trust boundaries, version skew, and heterogeneous operators.
Because registry controls do not accompany the artifact, end-to-end verification must be portable, making signing the primary defense against mirror and path attacks\cite{cappos2008look}.

\section{Discussions}

\subsection{Are Registries Really Security Hardened?}

While we have assumed that centralized registries are security-hardened and that software signing is primarily warranted because of additional distribution paths (mirrors, proxies, offline transfer), the reality is more complex. In practice, registry platforms and their surrounding infrastructure present multiple avenues for integrity failure within our threat model.

\paragraph{Registry break-ins and operator-side bugs}
Community package servers have been compromised before. In 2013, \texttt{RubyGems.org} was breached via a code-execution vulnerability in the web application stack, prompting a full service rebuild and gem verification~\cite{evan_2013}. In 2016, a server-side bug allowed potential replacement of certain uploaded \texttt{.gem} files; the RubyGems team issued mitigations and required verification by maintainers~\cite{Radcliffe_2016}. Similar risks also arise in adjacent infrastructure and redistribution layers (e.g., mirrors, enterprise caches, CDNs), where content may be stale, misconfigured, or maliciously served in the absence of end-to-end verification~\cite{cappos2008look,debian_secureapt_manual}.

\paragraph{A brief note on related attack-surface research}
Google Project Zero’s \emph{Windows Registry Adventure 7} study systematically maps the Windows Registry’s attack surface, showing how a single, central keystore accrues complex, widely exploitable behaviors when many components depend on it~\cite{Forshaw2025RegistryAdventure7}. Although the Windows Registry is not a software \emph{distribution} registry, the lesson is instructive: central control points attract attacks, and complexity creates room for subtle integrity failures. By analogy, software package registries and their redistribution layers (mirrors, proxies, CDNs) are high-value control surfaces that should not be assumed perfectly hardened.

\noindent
\paragraph{Zero Trust for Software Distribution}
In industry, \emph{Zero Trust} is a straightforward rule: never trust, always verify. Applied to software delivery, this means an organization should not place inherent or persistent trust in infrastructure it does not control, public registries, mirrors, proxies, and CDNs included. Registry checks are useful at download time, but verification must also occur at the point of use and across any redistribution. Given the operator errors, misconfigurations, and compromise avenues outlined above, a Zero Trust posture fits the risk: signatures and attestations keep identity and integrity verifiable wherever the artifact is obtained. Put differently, even if the forecast calls for clear skies (a “hardened” registry), bring an umbrella, signing, so assurance travels with the artifact, not the weather(registry).



\subsection{Trust Boundaries with AI Intermediaries}
The entry and proliferation of AI agents and large language models (LLMs) in software production introduces new considerations for the importance of software signing in supply chain security. In particular, AI systems add a new trust boundary: between the AI intermediary and the human or organizational entities that ultimately consume its outputs.

The recent integration of LLMs into coding workflows creates the possibility that these agents will suggest or automatically fetch software components that have been compromised, \eg Slopsquatting\cite{Gooding_2025}. This includes typosquatted packages, dependency confusion, or trojaned versions of legitimate artifacts. Moreover, AI can also be weaponized by attackers themselves to automate and scale these threats. A concrete example is the 2022 \texttt{solana-py} typosquat on PyPI, which masqueraded as the legitimate \texttt{solana} library but was designed to exfiltrate users’ cryptocurrency keys~\cite{Sonatype2022SolanaPy}. Although this particular incident was not generated by AI, it highlights the type of supply chain compromise that AI could make more prolific by automating large-scale typosquat-type attacks like these. In this way, the proliferation of AI lowers the barrier to entry for attackers, making such campaigns easier to launch, more frequent, and more difficult to detect.

In such cases, software signing provides two critical assurances. First, it enables automatically verifiable integrity checks that AI agents could enforce before including a component, thereby making the agent’s decisions auditable. Second, signing limits the attacker’s ability to impersonate trusted publishers, even in the face of massively scaled attack campaigns. Together, these properties ensure that the trust relationship between humans and AI intermediaries remains grounded in verifiable cryptographic guarantees rather than blind reliance on AI outputs.


\section{Conclusion}
\label{sec:conclusion}

Our position is straightforward.
Registry hardening improves security inside its boundary, but it does not make trust portable.
Software signing binds producer identity to artifact integrity so verification travels with the artifact across mirrors, proxies, CDNs, and air‑gapped transfers.
We therefore treat signing as a baseline layer of defense in modern software distribution.
This lens helps teams decide when registry-local controls are sufficient and when portable verification via signing is required.

\newpage

\bibliographystyle{IEEEtran}
\bibliography{bibliography/references, bibliography/new}


\end{document}